\documentclass[a4paper,conference]{IEEEtran} %
\usepackage{multirow}
\usepackage{cite} 
  \usepackage[pdftex]{graphicx}
  \usepackage[caption=false,font=footnotesize]{subfig}

\usepackage[cmex10]{amsmath} 
\usepackage{amssymb}
\usepackage{booktabs} 

\usepackage{tabu}
\usepackage{multirow}

\newcommand\T{\rule{0pt}{2.9ex}}       
\newcommand\B{\rule[-1.2ex]{0pt}{0pt}} 

\DeclareMathOperator*{\argmax}{arg\,max}

\begin{document}


\title{Leveraging Synthetic Subject Invariant EEG Signals for Zero Calibration BCI}




\author{\IEEEauthorblockN{Nik Khadijah Nik Aznan\IEEEauthorrefmark{1}, Amir Atapour-Abarghouei \IEEEauthorrefmark{2}, Stephen Bonner \IEEEauthorrefmark{1} \IEEEauthorrefmark{2},
\\ Jason D. Connolly\IEEEauthorrefmark{1} and Toby P. Breckon\IEEEauthorrefmark{1}}

\IEEEauthorrefmark{1} Durham University, Durham, UK $\|$
\IEEEauthorrefmark{2} Newcastle University, Newcastle, UK }

\maketitle

\begin{abstract}

  Recently, substantial progress has been made in the area of Brain-Computer Interface (BCI) using modern machine learning techniques to decode and interpret brain signals. While Electroencephalography (EEG) has provided a non-invasive method of interfacing with a human brain, the acquired data is often heavily subject and session dependent. This makes the seamless incorporation of such data into real-world applications intractable as the subject and session data variance can lead to long and tedious calibration requirements and cross-subject generalisation issues. Focusing on a Steady State Visual Evoked Potential (SSVEP) classification systems, we propose a novel means of generating highly-realistic synthetic EEG data invariant to any subject, session or other environmental conditions. Our approach, entitled the Subject Invariant SSVEP Generative Adversarial Network (SIS-GAN), produces synthetic EEG data from multiple SSVEP classes using a single network. Additionally, by taking advantage of a fixed-weight pre-trained subject classification network, we ensure that our generative model remains agnostic to subject-specific features and thus produces subject-invariant data that can be applied to new previously unseen subjects. Our extensive experimental evaluation demonstrates the efficacy of our synthetic data, leading to superior performance, with improvements of up to 16 percentage points in zero-calibration classification tasks when trained using our subject-invariant synthetic EEG signals.
\end{abstract}

\IEEEpeerreviewmaketitle

\section{Introduction}
\label{sec:intro}

Brain Computer Interface (BCI) can be an invaluable communication and control medium for people with severe physical disabilities, such as the Complete Locked-In Syndrome or Amyotrophic Lateral Sclerosis \cite{bi2013eeg}, as it can enable the use of various bespoke communication interfaces \cite{podmore2019relative} or robots for various tasks that are otherwise impossible due to such disabilities. If signals streamed from the human brain cortical are received and interpreted as commands, even real-time robotic navigation based on brain signals is practical \cite{Rao2013,sheng2017design}. 

In this work, we predominantly focus on the Steady State Visual Evoked Potential (SSVEP) paradigm as the visual stimuli, which evokes a neurophysiological response in a human viewing a frequency-based visual stimulus \cite{Rao2013}, making it a prime candidate for use in teleoperation tasks. Electroencephalography (EEG) is a prominent signal acquisition method in BCI  \cite{li2019multisource, corley2018deep,  Rao2013}, with the bio-signals obtained in a non-invasive, easily-decodable manner. The recently popularised dry electroencephalography can offer a commercial alternative that alleviates the cumbersome steps required for wet EEG, such as skin preparation and electrode gel application \cite{aznan2018classification}.

\begin{figure}[!t]
  \vskip -10pt
  \centering
  \includegraphics[width=0.65\linewidth]{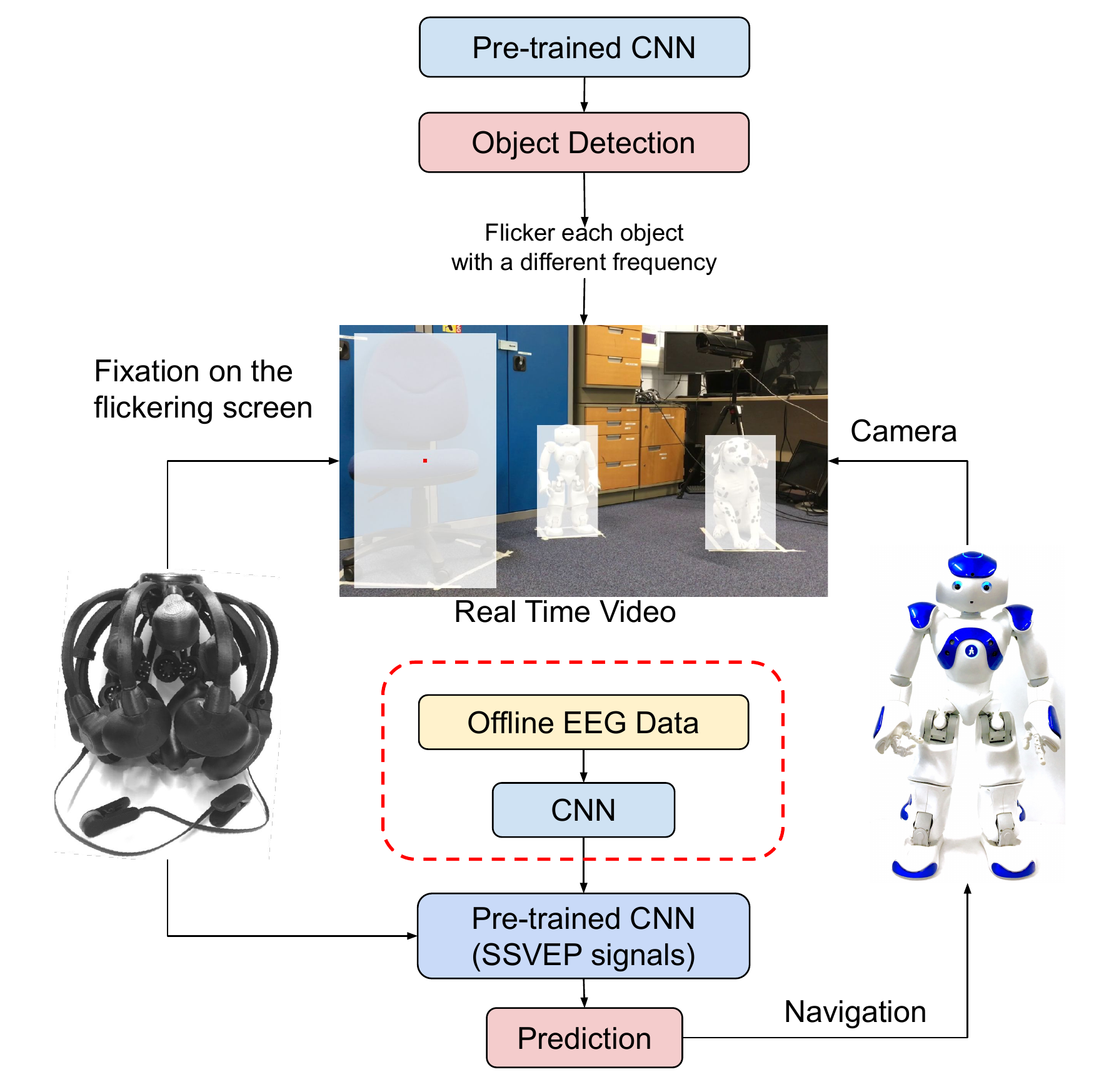} 
  \vskip -5pt
  \caption{Teleoperation task with the required calibration stage identified in the red dotted block.}
  \vskip -20pt
  \label{fig:overview}
  \end{figure} 

However, the resulting EEG signals are highly subject and session dependent, so much so that the signals contain patterns unique to specific subjects and have even been used as a biometric \cite{fazli2009subject, yu2019eeg}. This leads to severe calibration requirements for any EEG-based BCI application. Calibration is needed at the beginning of any task to account for specific subject and session conditions \cite{krauledat2008towards, lotte2018review} as illustrated in Figure \ref{fig:overview}.   

\begin{figure}[!h]
  \vskip -5pt
  \centering
  \includegraphics[width=0.65\linewidth]{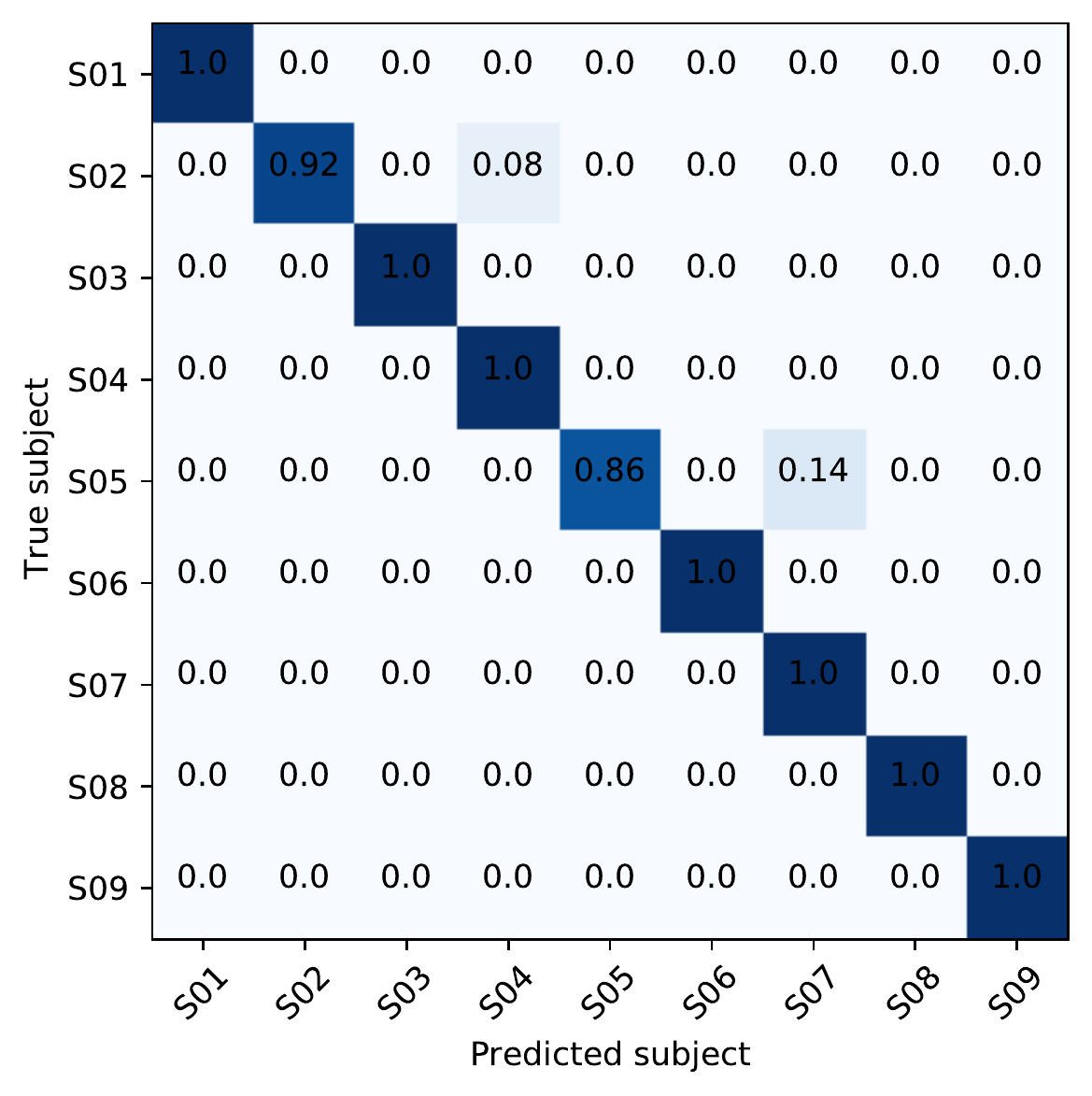} 
  \vskip -5pt
  \caption{Confusion matrix for subject classification accuracy (Maximal result being accuracy = 1.0 on the diagonal).}
  \vskip -5pt
  \label{fig:sub_cm}
\end{figure}

Many EEG-based applications, for example teleoperation, could be used as assistive technologies if not for their subject and session dependency \cite{lotte2018review}, that restricts real-world practical applications. This extreme subject dependence is demonstrated in Figure \ref{fig:sub_cm}, where a confusion matrix is seen for subject-biometric classification. A simple convolutional neural network easily manages to classify SSVEP-based EEG signals from nine subjects, indicating the existence of a distinctive subject bias, where each subject signal contains easily-distinguishable biometric features. In this paper, we investigate the possibility of leveraging recent advances in deep machine learning to generate synthetic EEG signals which are subject invariant and hence can be used to enable the training of generic EEG signal decoders for previously unseen subjects. Recently, neural-based generative models, such as Generative Adversarial Networks (GAN) \cite{goodfellow2014generative}, have been proven to be capable of capturing the key elements of a given dataset by learning a hidden structure from its underlying distribution to generate new data samples within the same distribution. Previous work \cite{aznan2019simulating} has even demonstrated significant improvements in performance if synthetic EEG signals generated by a GAN are used for pre-training an SSVEP classification network.

However, as EEG signals are highly subject-specific, the issue of subject bias has to be taken into consideration when dealing with such data in a learning-based approach \cite{kim2019learning, xu2018fairgan}. Machine learning, in general, is fraught with data-specific issues as models often have a tendency to overfit to unapparent or unobservable nuances of a given dataset in order to maximise task performance. Particular features in a dataset can often drive the resultant model distribution toward a certain direction resulting in bias and inferior performance on previously unseen data. 

In this paper, we exploit the capabilities of a Subject Invariant SSVEP Generative Adversarial Network (SIS-GAN) to produce not only realistic synthetic EEG signals but also to eliminate subject-specific features in order to boost performance on subsequent tasks. To demonstrate this, we choose to show results on an SSVEP-based teleoperation task, although potential applications are not limited just to this. In this cross-task setting, the data utilised to train our generative model are from different subjects and different capture sessions that remain unseen for the downstream SSVEP classifier. The generated signals are used to classify unseen subjects from the online EEG signals from \cite{aznan2019using}, where three subjects navigate a humanoid robot using variable position and size SSVEP stimuli. Our experimental results (Section \ref{sec:results}) demonstrate the efficacy of our SIS-GAN approach.

In summary, the major contributions of this paper are: 
\begin{itemize}
  \item \emph{Zero Calibration} - Our novel model architecture results in synthetically-generated subject-invariant EEG signals suitable for real-time BCI applications with near zero calibration.
  
  \item \emph{Cross-Task Results} - Downstream models trained on data generated by our SIS-GAN approach are shown to perform well when classifying Online signals having only seen Offline signals during training.
  
  \item \emph{Realistic Synthetic Data} - We use our deep learning based generative model to create highly \emph{realistic synthetic EEG signals} shown to offer superior performance for downstream signal decoding tasks.
  
  \item \emph{SSVEP Classification for Unseen Subjects} - Our use of enforced subject invariance, via our SIS-GAN architecture, results in models capable of accurate SSVEP classification for unseen subjects with improvements of up to 16 percentage points against contemporary state-of-the-art approaches.
  
\end{itemize}

Our extensive experimental results are fully supported by the release of our \emph{PyTorch} implementation\footnote{https://github.com/nikk-nikaznan/Subject-Invariant-SSVEP-GAN}.
\section{Related Work}
\label{sec:related-work}

Calibration issues have long plagued BCI applications, as calibration is required at the beginning of every session \cite{krauledat2008towards}, even for the same subjects doing the same tasks, leading to subject fatigue, and hence affecting signal quality. Removing or reducing the calibration stage, therefore, has received considerable attention within the literature. For instance, a zero training method is introduced in \cite{krauledat2008towards} to learn spatial patterns in features by transferring knowledge from previous sessions. Though effective in reducing task time, the approach is subject-specific and calibration is still required for new unseen subjects. In \cite{li2019multisource}, calibration requirements are somewhat reduced through style transfer mapping between previously seen and new subjects. Each new subject, however, will still require a few calibration sessions to enable training of the mapping process. Complete generalisation to entirely new subjects without calibration is, however, possible as demonstrated in our previous work \cite{aznan2018classification}, where an SSVEP-based task could be performed for previously unseen subjects without any additional training despite the slight reduction in accuracy.

In this paper, we propose a novel EEG signal generation framework taking advantage of generative models to produce subject-invariant synthetic EEG signals. However, the use of generative models in BCI applications is not unprecedented. In \cite{aznan2019simulating}, for instance, we investigated the possibility of generating synthetic EEG signals containing SSVEP information via neural-based generative models trained on a limited quantity of EEG signals from different subjects. Similarly, in \cite{hartmann2018eeg}, a Wasserstein GAN \cite{arjovsky2017wasserstein} is trained using single-channel EEG-based motor imagery data. High-resolution EEG-based motor imagery data is also generated in \cite{corley2018deep} via a similar model by interpolating one channel to another using low-resolution signals. Using synthetic signals generated from a Recurrent GAN, \cite{abdelfattah2018augmenting} augments motor imagery data with synthetic signals but without much improvement over the sole use of real data.

As subject invariance is an important component of our work, removing undesirable bias from data features is of significant relevance. In \cite{kim2019learning}, a model trained on biased data incentivised not to learn some specified target bias is shown to be able to perform well on unbiased test data. The model in \cite{xu2018fairgan} generates data without having any information about certain \emph{protected attributes} using a generator conditioned on those attributes and two discriminators discerning the fakeness of the samples and the existence of the undesirable features. 

Inspired by these advances in unbiased training frameworks, we investigate synthetic EEG signals generated via our Subject Invariant SSVEP Generative Adversarial Network (SIS-GAN) used to classify unseen subjects in a downstream EEG-based teleoperation task. While earlier work \cite{krauledat2008towards,li2019multisource,hwang2019ezsl,ozdenizci2020learning} discusses the difficulties often faced with the calibration stage, no other work to date has successfully managed to classify completely unseen subjects without re-training, fine-tuning or much larger data availability. In this vein, we propose a novel adversarial generative model (Section \ref{sec:SIGAN}) capable of producing highly valuable subject-invariant EEG signals that could be used to augment the training dataset needed for EEG-based models, leading to a significant boost in performance.  
\section{Proposed Approach}
\label{sec:approach}

In this section, we outline our proposed Subject Invariant SSVEP Generative Adversarial Network (SIS-GAN) approach to generating subject invariant EEG signals containing SSVEP information. The SIS-GAN model architecture is presented in Figure \ref{fig:SI-ACGAN} and comprises four primary components: a Generator network and its corresponding Discriminator (Section \ref{sec:DCGAN}), an Auxiliary classification network (Section \ref{sec:ACGAN}) and a pre-trained Subject-biometric classifier (Section \ref{sec:SIGAN}), enabling the generation of subject invariant signal samples.

\subsection{Generator and Discriminator Networks}
\label{sec:DCGAN}

The backbone of our approach for EEG signal generation is inspired by prior work on producing synthetic images \cite{radford2015unsupervised}. In this Generative Adversarial Network (GAN) setup, the generator, $G$, receives as its input random noise vectors, $z$, sampled from a Gaussian distribution. This generator produces fake data samples ($\tilde{x} = G(z)$) at every iteration, which along with real data samples, $x$, are used as the input to a discriminator $D$. As the discriminator is trained to classify the data samples as either fake or real, the resulting gradients are successively used to train the generator, leading to higher quality fake samples to the point where they become indistinguishable from the real ones. The training objective consequently relies on the competition between the generator and the discriminator following the minimax objective \cite{goodfellow2014generative}:

\vskip -10pt
\begin{equation}  
    L_D =\min_{G} \max_{D}\ \mathop{\mathbb{E}}_{x\sim \mathbb{P}_{r}} [log(D(x))] + \mathop{\mathbb{E}}_{\tilde{x} \sim \mathbb{P}_{g}} [log(1 - D(\tilde{x}))],
    \label{eq:vanilla-GAN}
\end{equation}
where $\mathbb{P}_{r}$ is the real data distribution, $\mathbb{P}_{g}$ the model distribution defined by $\tilde{x} = G(z), z \sim p(z)$, and $z$ the random noise vector used as the input to the generator.

It is important to note that a vanilla GAN would only be capable of capturing the underlying distribution of and thus generating one particular class of data at a time. Due to this limitation, in the following section, we detail the introduction of an auxiliary classification component to allow the model to produce data from multiple classes simultaneously. 

\subsection{Auxiliary Classifier Network}
\label{sec:ACGAN} 

An interesting alternative to training separate models that would generate data on a class by class basis is to have one model capable of producing data from all classes as required. In an Auxiliary Classifier GAN (AC-GAN), \cite{odena2017conditional}, the input to the generator is not only a random noise vector but also a class label for the generated output. The generator is consequently trained to produce fake data samples from a model distribution similar to the real data distribution for each specific class label. The discriminator will essentially identify whether the generated data is real or fake while at the same time classifying which class the generated data sample belongs to. Not unlike \cite{radford2015unsupervised}, the discriminator and the generator networks in an AC-GAN approach are trained to maximise each other's objective function but with two loss components: loss of the source (real or fake) and loss of the class label. As such, the use of an AC-GAN enables us to train a single model capable of generating EEG signals for all subjects and all three SSVEP classes. 

As the auxiliary component is fundamentally a classification task, cross-entropy is used as the loss function. Cross-entropy functions by measuring the probability between the predicted class ($x$) value to the actual value ($y$): 
\vskip -10pt
\begin{equation}  
  L_A = - \Sigma (x \log y).
  \label{eq:aux-loss}
\end{equation}
\vskip -10pt

\subsection{Creating Subject Invariant EEG Signals}
\label{sec:SIGAN}

Inspired by recent work on bias removal \cite{xu2018fairgan, kim2019learning}, we propose a novel synthetic EEG signal generation pipeline capable of producing subject-invariant signals that are unbiased representative samples for any unseen subject. Knowing the target bias, \cite{kim2019learning} proposes a network targeting specific features to penalise the model when said features become prominent in the model distribution. For instance, a network can easily be biased towards using colour information as a cue while colour has indeed no relevance to the actual task, such as handwritten digit classification. Therefore, by specifically targeting RGB information, the bias can be eliminated, leading to a more robust representation learning. However, in our dataset, there is no specific feature that can be identified as the culprit when it comes to introducing subject bias. EEG data is known to be complex, with every subject and session being uniquely identifiable - a phenomenon still not fully understood \cite{krauledat2008towards, li2019multisource}.

\begin{figure}[!t]
  \centering
  \includegraphics[width=0.78\linewidth]{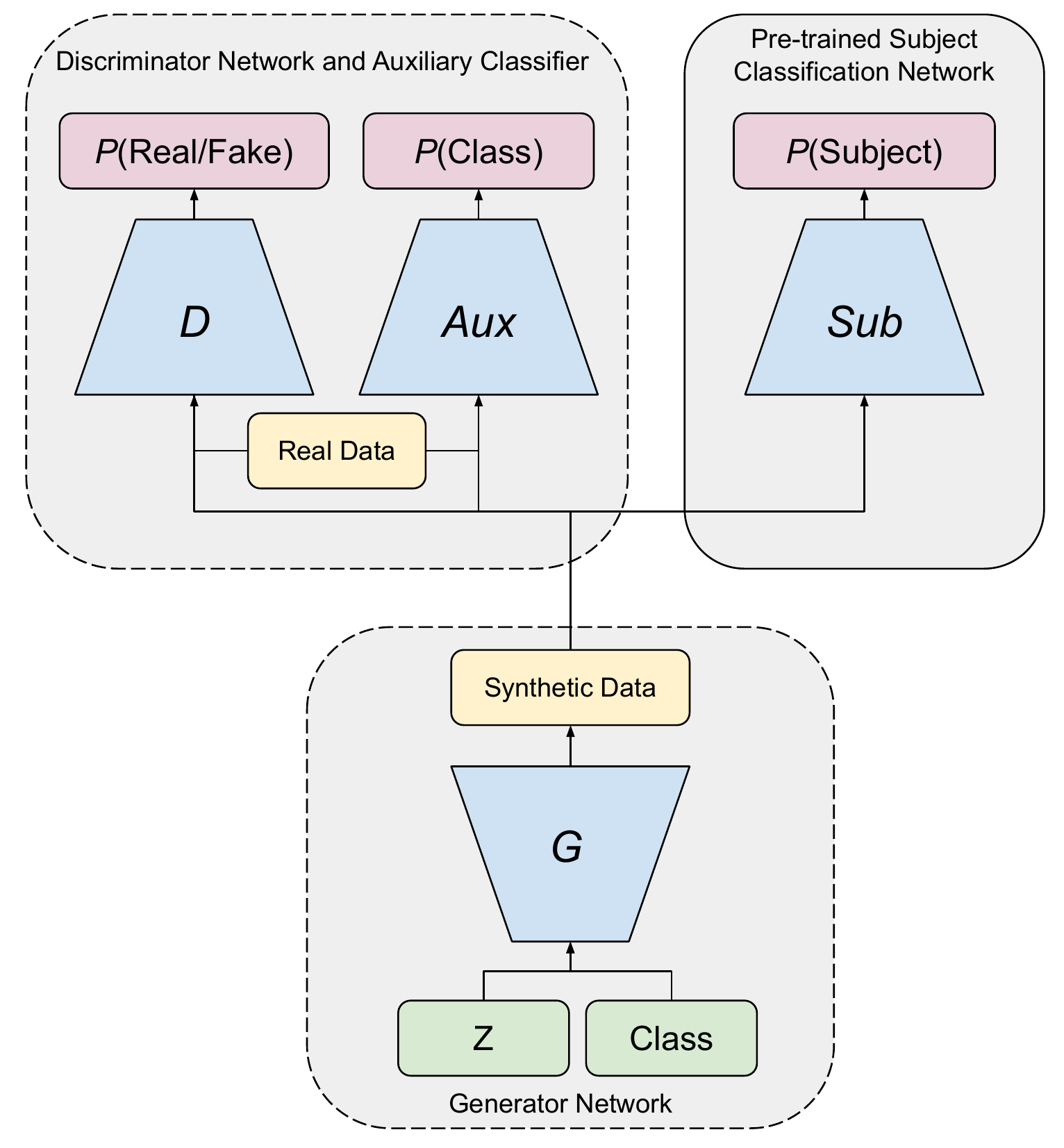} 
  \vskip -5pt
  \caption{Our Subject Invariant SSVEP Generative Adversarial Network. The Generator (G) produces data with subject-specific information removed (Sub) that can fool the Discriminator (D) and is classified as a certain frequency (Aux).}
  \label{fig:SI-ACGAN}
  \vskip -20pt
\end{figure}

Consequently, we design our training process in such a manner that the network is rewarded when it learns features that are common across all subjects and is penalised when subject-specific features can lead to subject-biometric classification. To achieve this, we introduce a pre-trained frozen subject-biometric classification network (shown in the upper-right of Figure \ref{fig:SI-ACGAN}), charged with classifying which subject the generated EEG signal biometrically belongs to. It is important to reiterate that this classification network is frozen and the gradients from its loss function are only used to train the generator and not the network itself. Since a correct classification would be undesirable, the gradients from this pre-trained frozen network can be used to penalise the generator, thus pushing it towards generating more generic subject-invariant outputs. More formally, the subject invariant loss component can be viewed as the following objective:

\vskip -10pt
\begin{equation}
L_S = \argmax_{\hat{y}} S(\hat{y}|x),  
\label{eq:sub-loss} 
\end{equation}
where $S$ is the pre-trained frozen subject-biometric classification network, $x$ denotes the generated data and $\hat{y}$ is the predicted probability values for all subjects. Essentially, this component minimises the maximum predicted subject probability which, over multiple training steps, will produce synthetic EEG signals which cannot be correctly classified by the subject network -- thus making the data subject invariant. 

Figure \ref{fig:SI-ACGAN} illustrates the overall SIS-GAN architecture and shows how the various components of our approach are connected. The model is able to generate samples for all classes using a single generator, but is similarly capable of producing subject-invariant EEG signals via the penalty introduced from the subject-biometric classification network. The overall training objective of our model is as follows:
\vskip -10pt
\begin{equation}
L = L_D + \lambda_a L_A + \lambda_s L_S,
\end{equation}
where $\lambda_a$ and $\lambda_s$ are used to weight the importance of the auxiliary classification and subject identification components in the overall loss score.

\subsection{Implementation Details}
\label{sec:Implementation}

For the sake of consistency, all the discriminator and classification networks in Sections \ref{sec:DCGAN}, \ref{sec:ACGAN} and \ref{sec:SIGAN} follow a similar architecture with four layers containing modules of 1D Convolution, BatchNorm, PReLU and DropOut ($p$ = 0.5) followed by a linear layer projecting the resulting features to the number of desired classes. The discriminator in AC-GAN (Section \ref{sec:ACGAN}) includes two heads, one for discriminating between real and fake samples and one for auxiliary classification. No max-pooling is used as our experiments show strided convolutions yield better performance.

The architecture of the generator in all models contains five layers of fractionally-strided convolutions with BatchNorm and PReLU. Our experiments with residual connections \cite{he2016deep} led to no significant improvements in the results. All implementation is done in PyTorch \cite{pytorch}, with Adam \cite{kingma2014adam} providing the optimisation ($\beta_{1} = 0.5$, $\beta_{2} = 0.999$, $\alpha = 0.0001$). In SIS-GAN (Section \ref{sec:SIGAN}), the subject loss in Equation \ref{eq:sub-loss} is empirically weighted down (by a factor of $0.3$ for $\lambda_s$) for a more stable training process and improved results.

\section{Experimental Setup}
\label{sec:experimental-setup} 

\subsection{Datasets}
For our experiments, we make use of two sets of Dry-EEG data from non-overlapping subjects, collected under two different conditions: Offline and Online. Both datasets are recorded using Quick-20 dry EEG headset from Cognionics Inc. with 20 dry-EEG sensors. We collect data over the parietal and occipital cortex (P7, P3, Pz, P4, P8, O1 and O2) \cite{Lin2014Assessing}, frontal center (Fz) and A2 reference at 500 Hz sample rate with the stimuli displayed on a 60Hz LCD monitor.

\subsubsection{Offline Real Dry-EEG Dataset}
\label{subsec:Offline} 

Data is collected from nine inexperienced subjects (Na\"{i}ve BCI users) aged 25 to 40 years old from an offline SSVEP experiment (S01 to S09). Cortical signals from the subjects are streamed and recorded using the dry-EEG headset whilst their gaze is fixated on variable SSVEP stimuli from \cite{aznan2019using}. The stimuli is created using detected objects in a video sequence. Black and white boxes are rendered over object blocks, thus simulating a flickering effect with display frequency modulations of 10, 12 and 15 Hz to create SSVEP frequencies. Alternately, the interface displays navigational arrows with the same frequency modulations. The objects in the video sequence are captured using the camera onboard our humanoid robot, NAO \cite{gouaillier2009mechatronic}. We collect 60 samples for each of the three classes per subject.

\subsubsection{Online Teleoperation Task Dry-EEG Dataset}
\label{subsec:Online} 

Data is collected using our experimental setup for real-time robotic teleoperation \cite{aznan2019using} based on three subjects (T01, T02, T03). No subject from the Offline dataset is used and all subjects are experienced participants. Subjects are seated in front of a computer screen to navigate a humanoid robot by fixating on real-time on-screen stimuli. The robot faces a scene containing objects which are detected and flickered with different unique frequencies. The decoded SSVEP signals are then used to navigate the humanoid robot towards the subject-selected objects within the calculated robot motion trajectory. The stimuli interface displayed to the subject alternates between flickering objects and the navigational arrows to allow the robot to be navigated when there are no new objects detected within the scene. To enable the use of this data, the raw EEG signals and the ground truth information are all saved from the real-time experiment. The data contains 30 unique samples per subject. We use this dataset in the cross-task validation experiment (Section \ref{sec:unseen}), where the objective is to learn a model from the Offline data such that it can correctly classify the data in the Online dataset. From the real-time task presented in \cite{aznan2019using}, all subjects demonstrate strong statistical real-time performance based on offline training, with the mean accuracy of 90, 87 and 80$\%$ respectively.

\subsection{Evaluation Methodology}
\label{sec:EM}
 
The main purpose of this work is to investigate the capability of a generative model to produce realistic synthetic signals that can potentially be used in real-time SSVEP classification to eliminate the calibration stage of a BCI application.

\begin{figure}[!h]
  \vskip -5pt
  \centering
  \includegraphics[width=0.72\linewidth]{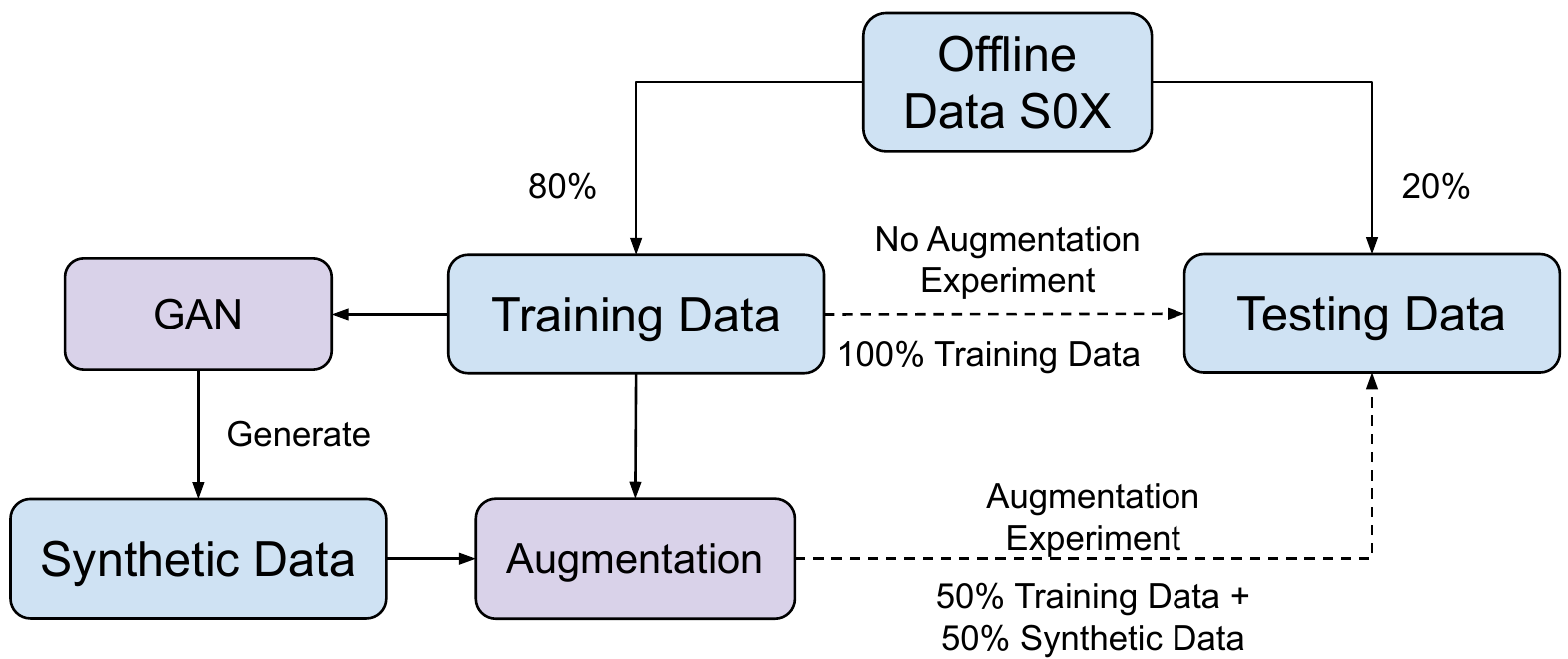} 
  \caption{Evaluation method for SSVEP classification for single subject, where the training and testing data are from the same subject. Dotted lines indicate SSVEP classification.}
  \label{fig:persub}
  \vskip -15pt
\end{figure} 

The first set of experiments (Section \ref{sec:signals_class}) focus on SSVEP classification for a single subject. Using the Offline dataset, we test the SSVEP classification performance for each subject when the network is trained and tested on the same subject. Results are compared against models trained using realistic synthetic data generated by a Deep Convolutional GAN (DC-GAN) and an AC-GAN. The generative model is evaluated via the augmented dataset containing both real and synthetic data used to classify the testing dataset. Figure \ref{fig:persub} outlines the experimental procedure. We utilise 20$\%$ of the data from Section \ref{subsec:Offline} for testing. The training dataset is used for training the generative and the SSVEP classification models. 

There are two main experiments within this setup: firstly a baseline to see how accurately each subject performs the task and secondly to measure any improvements in performance by augmenting the training dataset with synthetic EEG signals. This is important as most of the time, na\"{i}ve BCI users do not perform as well as experienced users in a variety of BCI tasks and hence some inter-subject variation is to be expected \cite{renton2019optimising}.

It is commonly known that EEG signals have unique patterns containing specific subject information that leads to difficulty in unseen subject classification \cite{yu2019eeg}. The rest of our experiments focus on the ability of the generated synthetic data to improve the generalisation of the classification model given that there is no prior training on a particular subject. To rigorously evaluate our approach, experiments are carried out using three training datasets:

\begin{itemize}
  \item \emph{Real Training Data} - where the resultant model is trained solely on the real data from the Offline dataset with no synthetic component.
  \item \emph{Augmented Training Data} - where model training is performed using real data from the Offline dataset mixed with synthetic data from the generative models at a ratio of 50:50.
  \item \emph{Synthetic Training Data} - where model training is performed using only synthetic data from the generative models, with the same number of total samples as the \emph{Augmented Data}. 
\end{itemize}

\begin{figure}[!t]
  \centering
  \subfloat[Evaluation method for AC-GAN.]{\includegraphics[width=0.42\textwidth]{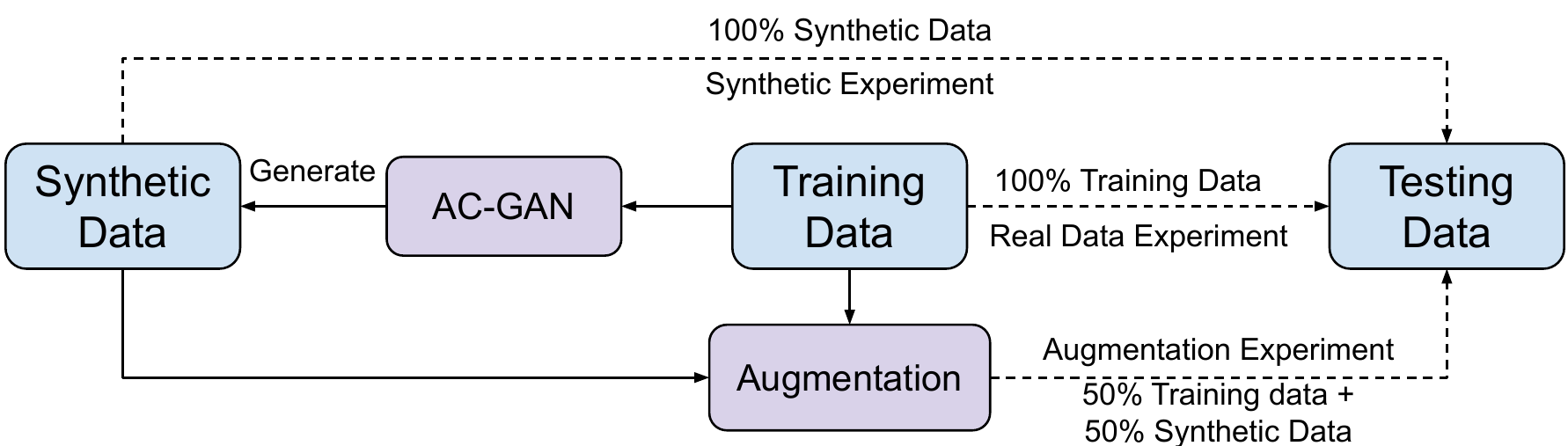}%
  \label{SSVEP Signal at 10Hz.}}
  \hfil
  \vskip -2pt
  \subfloat[Evaluation method for SIS-GAN.]{\includegraphics[width=0.42\textwidth]{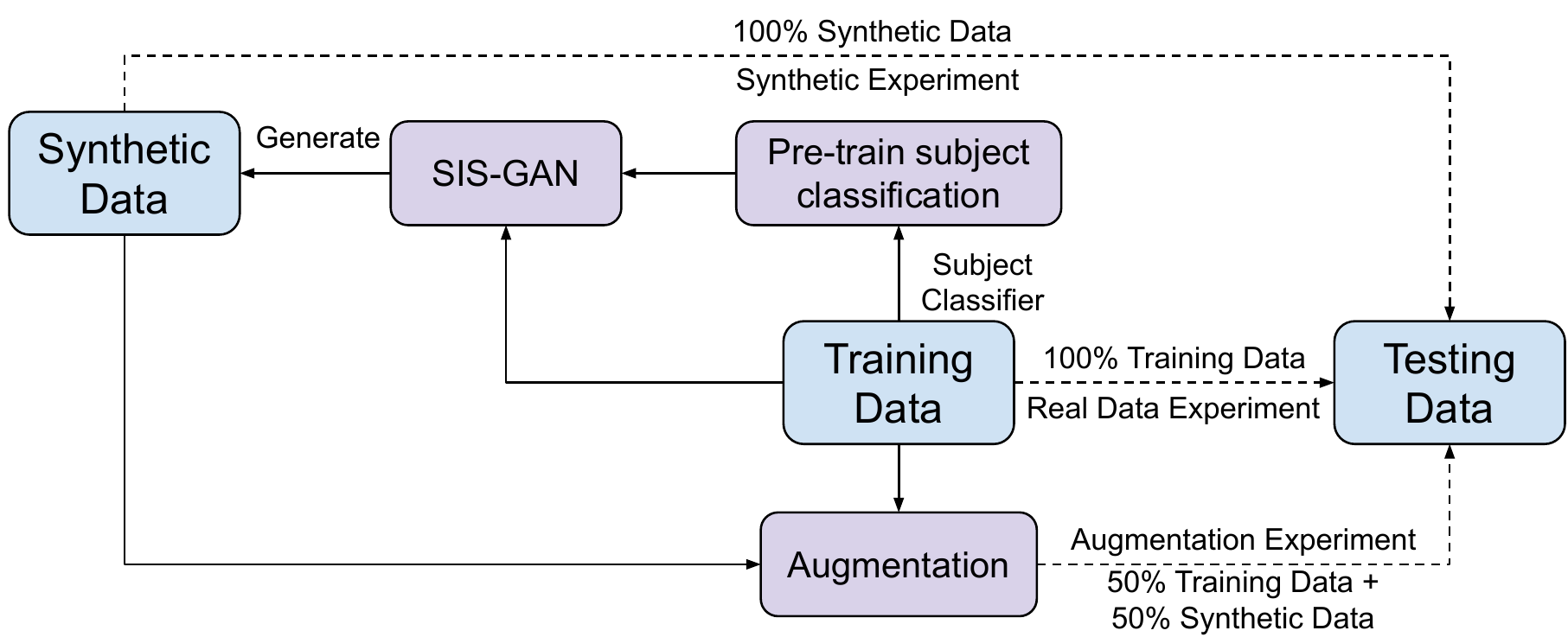}%
  \label{SSVEP Signal at 12Hz.}}
  \vskip -5pt
  \caption{Evaluation method for SSVEP classification for unseen subject where the test data is from an unseen subject with no prior training. Frozen subject-biometric classifier is used to train SIS-GAN. Dotted lines indicate SSVEP classification.}
  \label{fig:Unseen_EM}
  \vskip -15pt
\end{figure}

The second set of experiments (SSVEP Classification for Unseen Subject : Leave-One-Out -- Section \ref{sec:LOO}) likewise focus on the same Offline dataset $\{$D$\}$. However, instead of training and testing on data captured from one subject, we perform tests on data from one unseen subject $\{$S0x$\}$ after training the model on data from all other subjects $\{$D - S0x$\}$. Figure \ref{fig:Unseen_EM} shows the experimental procedure, where we are taking the leave-one-out validation approach (i.e. the leave-one-out is $\{$S0x$\}$ (test data) and data from all remaining subjects are used as the training data $\{$D - S0x$\}$).

Typically in prior works \cite{lotte2018review, aznan2019using}, real-time  SSVEP classification results are obtained by calibrating the SSVEP classifier model to the subject during the preliminary task. In our last set of experiments (SSVEP Classification for Unseen Subject : Cross-Task -- Section \ref{sec:unseen}), which are the primary focus of this work, we explore whether this calibration stage can be removed, thus requiring no model retraining before a subject participates in the session. To achieve this, we evaluate the performance of a model on unseen subjects (online task) when the model is pre-trained on data from a completely different task (offline task). Normally, to do this, we would ideally want as large a volume of data from different subjects as possible. However, this is far from ideal as data collection is both time-consuming and expensive \cite{abdelfattah2018augmenting}, further emphasising the importance of our approach. In our approach, we generate realistic data from a modest dataset containing data from nine subjects (detailed in Section \ref{subsec:Offline}). Different training data are used to train a SSVEP classifier model, which is able to accurately classify data from the three unseen subjects from the Online dataset (detailed in Section \ref{subsec:Online}). The procedure of this evaluation is shown in Figure \ref{fig:Unseen_EM}, where the SSVEP classifier model is trained using the Offline dataset and tested on the Online dataset.

\section{Experimental Results}
\label{sec:results}

The experimental results are presented here, with every result being the mean and standard deviation of ten different runs for each experiment, with a unique random seed used for every run. All experiments are performed using the same number of data points, with the augmented training dataset consisting of a mix of synthetic and real data at a ratio of 50:50. Final test experiments are performed on the real data. 

\subsection{SSVEP Classification for a Single Subject} 
\label{sec:signals_class}

We have shown in our previous work \cite{aznan2019simulating} that it is possible to generate realistic synthetic EEG signals containing SSVEP frequency features. The primary constraint in generating such synthetic EEG signals is the limited quantity of the real data available for training a model capable of generating synthetic data in the first place, as large datasets are required to train accurate generative models. Since collecting EEG signals is time-consuming and expensive, we explore incorporating an auxiliary classifier into our approach for generating synthetic data as the auxiliary classification component essentially exposes the model to three times the amount of data provided to a vanilla approach such as DC-GAN.

\begin{table}[h!]
  \centering
  \caption{Results of SSVEP classification on a single subject.}
  \label{table:persubject_result}
  \resizebox{\linewidth}{!}{

  \begin{tabu}{@{\extracolsep{3pt}}c c c c c c@{}}
  \hline\hline
  \multirow{2}{*}{\textbf{Subject}} & \textbf{No augmentation} & \multicolumn{2}{c}{\textbf{With augmentation}}\T\B\\
  \cline{2-2} \cline{3-4}
  & \textbf{Real Data} & \textbf{DC-GAN} & \textbf{AC-GAN}\T\B\\
  \hline\hline
  \textbf{S01} & 0.727 $\pm$ 0.069 & 0.582 $\pm$ 0.089 & \textbf{0.762 $\pm$ 0.036}\T\\
  \textbf{S02} & 0.843 $\pm$ 0.059 & 0.750 $\pm$ 0.089 & \textbf{0.868 $\pm$ 0.020}\\
  \textbf{S03} & \textbf{0.812 $\pm$ 0.048} & 0.732 $\pm$ 0.058 & 0.780 $\pm$ 0.027\\
  \textbf{S04} & 0.783 $\pm$ 0.073 & 0.533 $\pm$ 0.087 & \textbf{0.795 $\pm$ 0.027}\\
  \textbf{S05} & 0.807 $\pm$ 0.049 & 0.742 $\pm$ 0.076 & \textbf{0.828 $\pm$ 0.023}\\
  \textbf{S06} & 0.948 $\pm$ 0.024 & 0.860 $\pm$ 0.080 & \textbf{0.953 $\pm$ 0.016}\\
  \textbf{S07} & 0.962 $\pm$ 0.037 & 0.982 $\pm$ 0.029 & \textbf{0.995 $\pm$ 0.008}\\
  \textbf{S08} & \textbf{0.738 $\pm$ 0.086} & 0.500 $\pm$ 0.113 & 0.708 $\pm$ 0.033\\
  \textbf{S09} & 0.863 $\pm$ 0.058 & 0.860 $\pm$ 0.032 & \textbf{0.880 $\pm$ 0.025}\B\\
  \hline
  \textbf{Mean} & 0.831 $\pm$ 0.098 & 0.727 $\pm$ 0.172 & \textbf{0.841 $\pm$ 0.091}\T\B\\
  \hline\hline
  \end{tabu}

  }

\end{table}

\begin{table*}[!t]

  \caption{Mean accuracy with standard deviation when classifying SSVEP for unseen subject on offline dataset.}
  \label{table:unseen_result_loo}

  \centering
  \resizebox{0.75\linewidth}{!}{

  \begin{tabu}{@{\extracolsep{6pt}}c c c c c c@{}}
  \hline\hline

  \multirow{2}{*}{\textbf{Subject}} & \multirow{2}{*}{\textbf{Real data}} & \multicolumn{2}{c}{\textbf{Augmentated Training Data}} & \multicolumn{2}{c}{\textbf{Synthetic Training Data}}\T\B\\

  \cline{3-4} \cline{5-6}

  & &  \textbf{AC-GAN} & \textbf{SIS-GAN} & \textbf{AC-GAN} & \textbf{SIS-GAN}\T\B\\
  
  \hline\hline

  \textbf{Unseen S01} & 0.576 $\pm$ 0.055 &  0.534 $\pm$ 0.042  &  0.550 $\pm$ 0.028  &  0.617 $\pm$ 0.023  & \textbf{0.665 $\pm$ 0.017}\T\\
  \textbf{Unseen S02} & 0.705 $\pm$ 0.009 & 0.711 $\pm$ 0.023 & 0.709 $\pm$ 0.032 & 0.619 $\pm$ 0.053 & \textbf{0.714 $\pm$ 0.009}\\
  \textbf{Unseen S03} & 0.687 $\pm$ 0.016 & 0.688 $\pm$ 0.032 & \textbf{0.732 $\pm$ 0.021} & 0.604 $\pm$ 0.025 & 0.717 $\pm$ 0.006\\
  \textbf{Unseen S04} & 0.656 $\pm$ 0.010 & 0.581 $\pm$ 0.023 & 0.600 $\pm$ 0.028 & 0.627 $\pm$ 0.162 & \textbf{0.667 $\pm$ 0.012}\\
  \textbf{Unseen S05} & 0.682 $\pm$ 0.027 & 0.694 $\pm$ 0.024 & 0.711 $\pm$ 0.026 & 0.380 $\pm$ 0.029 & \textbf{0.768 $\pm$ 0.080}\\
  \textbf{Unseen S06} & 0.757 $\pm$ 0.045 & 0.710 $\pm$ 0.068 & 0.727 $\pm$ 0.032 & 0.627 $\pm$ 0.162 & \textbf{0.815 $\pm$ 0.029}\\
  \textbf{Unseen S07} & 0.905 $\pm$ 0.005 & 0.940 $\pm$ 0.009 & \textbf{0.943 $\pm$ 0.014} & 0.588 $\pm$ 0.155 & 0.911 $\pm$ 0.025\\
  \textbf{Unseen S08} & 0.447 $\pm$ 0.046 & 0.426 $\pm$ 0.038 & 0.378 $\pm$ 0.046 & 0.343 $\pm$ 0.030 & \textbf{0.485 $\pm$ 0.018}\\
  \textbf{Unseen S09} & 0.778 $\pm$ 0.012 & 0.807 $\pm$ 0.018 & \textbf{0.814 $\pm$ 0.023} & 0.710 $\pm$ 0.048 & 0.780 $\pm$ 0.009\B\\
  \hline
  \textbf{Mean} & 0.688 $\pm$ 0.125  & 0.677 $\pm$ 0.146 & 0.685 $\pm$ 0.155 & 0.543 $\pm$ 0.150 & \textbf{0.725 $\pm$ 0.113} \T\B\\
  \hline\hline

  \end{tabu}

  }
  \vskip -10pt
  \end{table*}

For this experiment, we attempt to classify SSVEP signals using data from a single subject -- this means that a new model is trained for each of the nine subjects. The results from this experiment are presented in Table \ref{table:persubject_result} and show that we are able to use synthetic data without losing much SSVEP classification performance. Additionally, the results show that there is a possibility that by augmenting synthetic data with real data, the classification result can actually be improved to some degree (from 83.1\% for training on real data alone to 84.1\% with data augmented by AC-GAN generated data). This can be seen as validation that realistic synthetic SSVEP data can indeed be generated. Based on the results presented here, and for the sake of brevity, for the remaining experiments we compare our results with AC-GAN as it has demonstrated the best performance on this task.



\subsection{Subject-biometric Classification} 
\label{sec:sub_class}
To evaluate our hypothesis that SSVEP-based EEG signals can be classified in accordance with the subject-biometric, we train a convolutional model (similar to \cite{aznan2018classification}), where the subject is used as the training label, rather than the SSVEP frequency. A confusion matrix detailing the performance of the model on each of the nine subjects from the Offline dataset is seen in Figure \ref{fig:sub_cm}. In addition, after using 10-fold cross validation, the model is able to produce a final mean test accuracy score over the nine subjects of 0.980$\pm$0.050. This demonstrates that EEG signals contain within them enough unique information which is accurately able to identify the subject from which the data was recorded. This result also corroborates with other research showing subject information to have a large detrimental impact on model performance when performing other tasks \cite{li2019multisource,hwang2019ezsl}. This is what primarily motivates the remainder of this work, as we try to eliminate these harmful subject-specific features when performing SSVEP classification. 

\subsection{SSVEP Classification for Unseen Subject : Leave-One-Out} 
\label{sec:LOO}

The pattern of electrical voltage produced by the brain often differs from one subject to the next, resulting in difficulty for a model to correctly classify SSVEP signals from a subject whose signals are absent from \textit{a priori} model training. To overcome such difficulty, we train a generative model to eliminate subject-specific features, potentially to be used on the unseen subject problem. For this experiment, we train both AC-GAN and SIS-GAN and evaluate the performance following the procedure outlined in Figure \ref{fig:Unseen_EM} in Section \ref{sec:EM}.


The results in Table \ref{table:persubject_result} demonstrate how training on real data can be effective when the training and test data are from the same subject, with mean accuracy of 83.1$\%$ (Table \ref{table:persubject_result} -- Real Data), but when the settings of the experiment change to testing on previously unseen test subject, using real data alone for training is not effective, as the mean accuracy is reduced to 68.8$\%$ (Table \ref{table:unseen_result_loo} -- Real Data). However, the results indicate how SIS-GAN can be a powerful tool to enhance unseen subject classification by generating subject-invariant synthetic data which can be used for training. As seen in Table \ref{table:unseen_result_loo}, when the model is trained only on synthetic EEG signals generated by our SIS-GAN model and tested on real data, an accuracy level of up to 72.5\% can be achieved, which is significantly higher than what a model trained on the same number of real data points can achieve (68.8\%).


\subsection{SSVEP Classification for Unseen Subject : Cross-Task} 
\label{sec:unseen}


\begin{figure*}[!t]
  \centering
  \subfloat[SSVEP Signal at 10Hz.]{\includegraphics[width=0.32\textwidth]{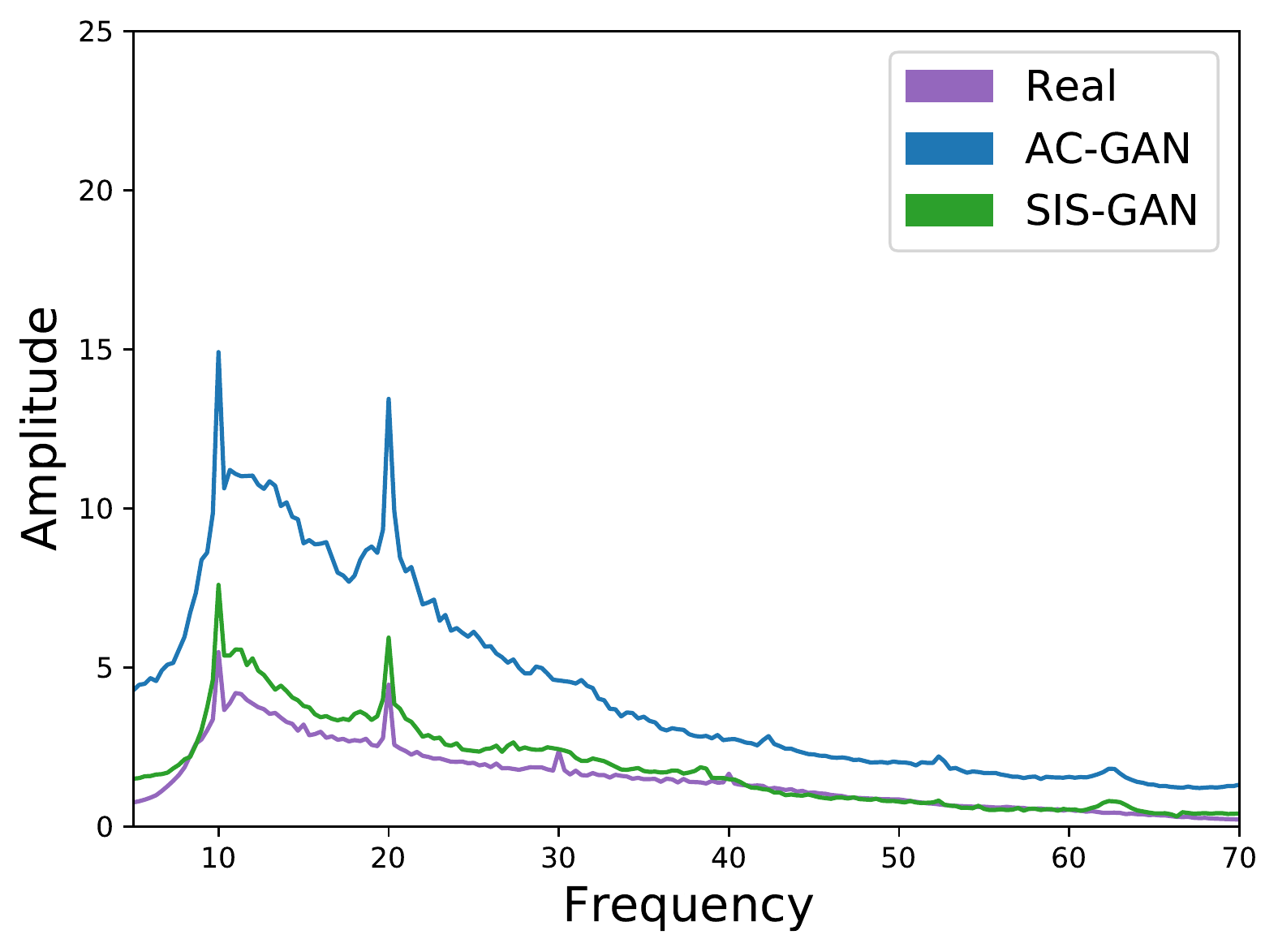}%
  \label{SSVEP Signal at 10Hz.}}
  \hfil
  \subfloat[SSVEP Signal at 12Hz.]{\includegraphics[width=0.32\textwidth]{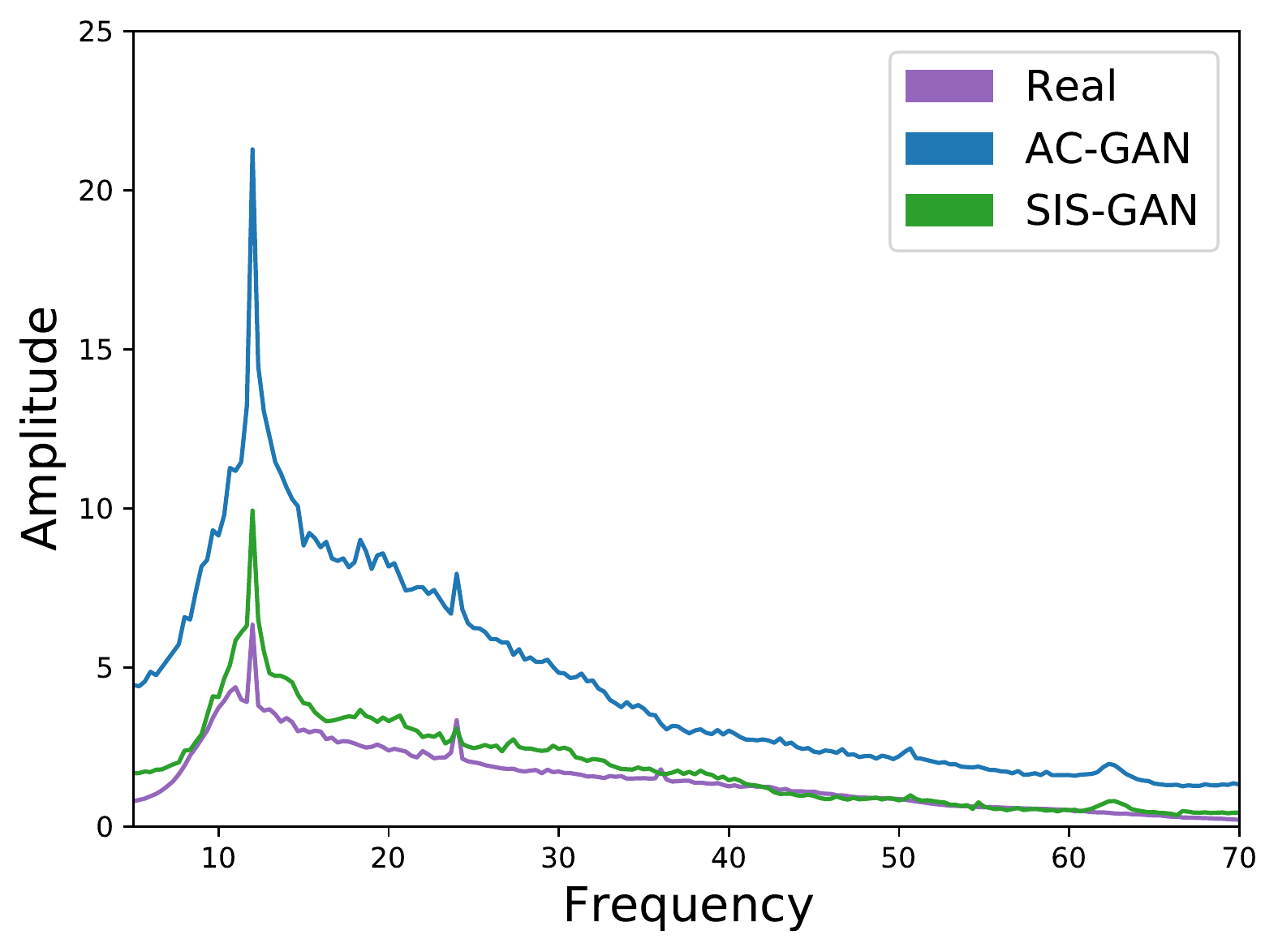}%
  \label{SSVEP Signal at 12Hz.}}
  \hfil
  \subfloat[SSVEP Signal at 15Hz.]{\includegraphics[width=0.32\textwidth]{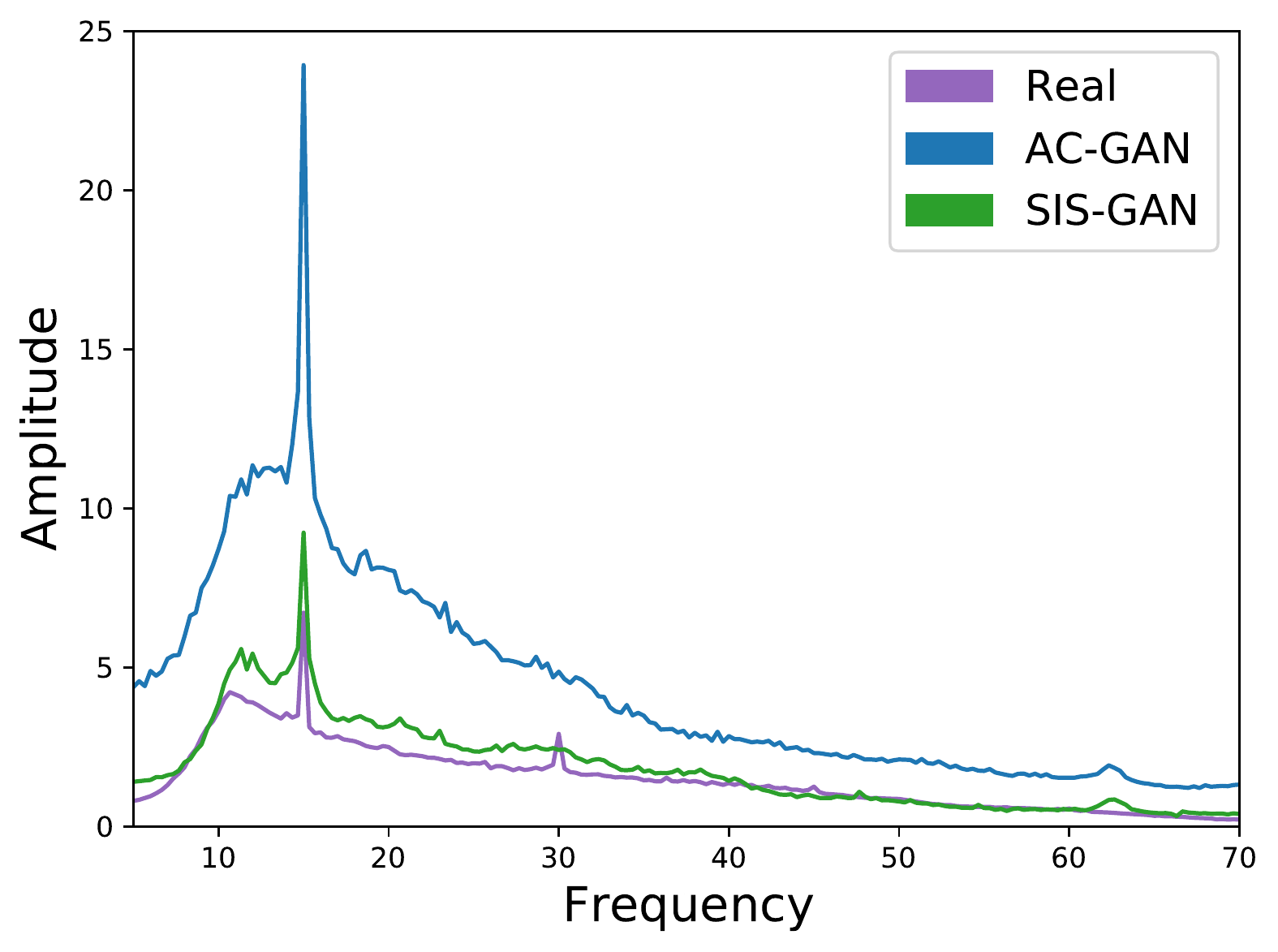}%
  \label{SSVEP Signal at 15Hz.}}
  \caption{Comparing Fast-Fourier Transforms (FFT) of real and synthetic data from all the generative models. The Synthetic data clearly displays the characteristic SSVEP frequency peaks and associated harmonics.}
  \vskip -5pt
  \label{fig:FFT}
\end{figure*}

We demonstrate that our proposed generative model is able to produce realistic EEG signals containing SSVEP frequency information. To achieve this, we analyse a subset of the generated data, taken from a model trained on the Offline dataset, using a Fast-Fourier Transform (FFT). The FFT plots for the three SSVEP frequencies taken from SIS-GAN and AC-GAN, as well as the real data, are presented in Figure \ref{fig:FFT}. The figure shows that when compared to the FFT taken from the real EEG signals, the generative models are capable of producing data which display the characteristic peak at the desired SSVEP frequency. It is interesting to note that although containing the correct peak, the data generated by AC-GAN often has a higher amplitude when compared to the real data.

\begin{table*}[!h]
  \caption{Mean accuracy with standard deviation for cross-task when classifying SSVEP on online subject dataset.}  
  \label{table:unseen_result_CT}

  \centering
  \resizebox{0.70\linewidth}{!}{

  \begin{tabu}{@{\extracolsep{6pt}}c c c c c c@{}}
  \hline\hline

  \multirow{2}{*}{\textbf{Subject}} & \multirow{2}{*}{\textbf{Real data}} & \multicolumn{2}{c}{\textbf{Augmentated Training Data}} & \multicolumn{2}{c}{\textbf{Synthetic Training Data}}\T\B\\

  \cline{3-4} \cline{5-6}

  & &  \textbf{AC-GAN} & \textbf{SIS-GAN} & \textbf{AC-GAN} & \textbf{SIS-GAN}\T\B\\
  
  \hline\hline

  \textbf{T01} & 0.736 $\pm$ 0.066 & 0.742 $\pm$ 0.071 & 0.756 $\pm$ 0.103 & 0.695 $\pm$ 0.100 & \textbf{0.845 $\pm$ 0.019}\T\\
  \textbf{T02} & 0.508 $\pm$ 0.072 & 0.542 $\pm$ 0.062 & 0.592 $\pm$ 0.052 & 0.488 $\pm$ 0.083 & \textbf{0.675 $\pm$ 0.016}\\
  \textbf{T03} & 0.240 $\pm$ 0.039 & 0.277 $\pm$ 0.067 & 0.303 $\pm$ 0.057 & 0.310 $\pm$ 0.060 & \textbf{0.453 $\pm$ 0.017}\B\\
  \hline
  \textbf{Mean} & 0.495 $\pm$ 0.211 & 0.520 $\pm$ 0.202 & 0.550 $\pm$ 0.201 & 0.498 $\pm$ 0.177 & \textbf{0.660 $\pm$ 0.161}\T\B\\
  \hline\hline

  \end{tabu}

  }
  \vskip -15pt
  \end{table*}

To illustrate the potential of our generative models, we perform cross-task classification as in Section \ref{sec:EM}. Cross-task classification requires a model trained solely on a dataset focusing on one task to generalise well to another dataset focusing on another task. In our experimental setup, we evaluate the performance of a model trained on the Offline dataset when tested using the Online dataset. This is a common real-world problem as often BCI systems are trained beforehand on existing data and expected to perform on different subjects in real time on a range of tasks including teleoperation \cite{aznan2019using}.

The experimental results are presented in Table \ref{table:unseen_result_CT}. One of the most striking observations that can be made from the results is that almost all of the approaches are an improvement upon just using the real data alone -- bringing some evidence that using our synthetically generated data can improve cross-task performance. Perhaps the most interesting result is that training a classification model only with data generated using SIS-GAN can lead to the model outperforming a model trained on the same number of real data points by over 16 percentage points. It is also of interest to observe that using the synthetic training data alone, SIS-GAN significantly outperforms AC-GAN, highlighting the potential benefits associated with removing subject-specific features from EEG signals. 

\begin{figure}[!t]
  \vskip -10pt
  \centering
  \includegraphics[width=0.75\linewidth]{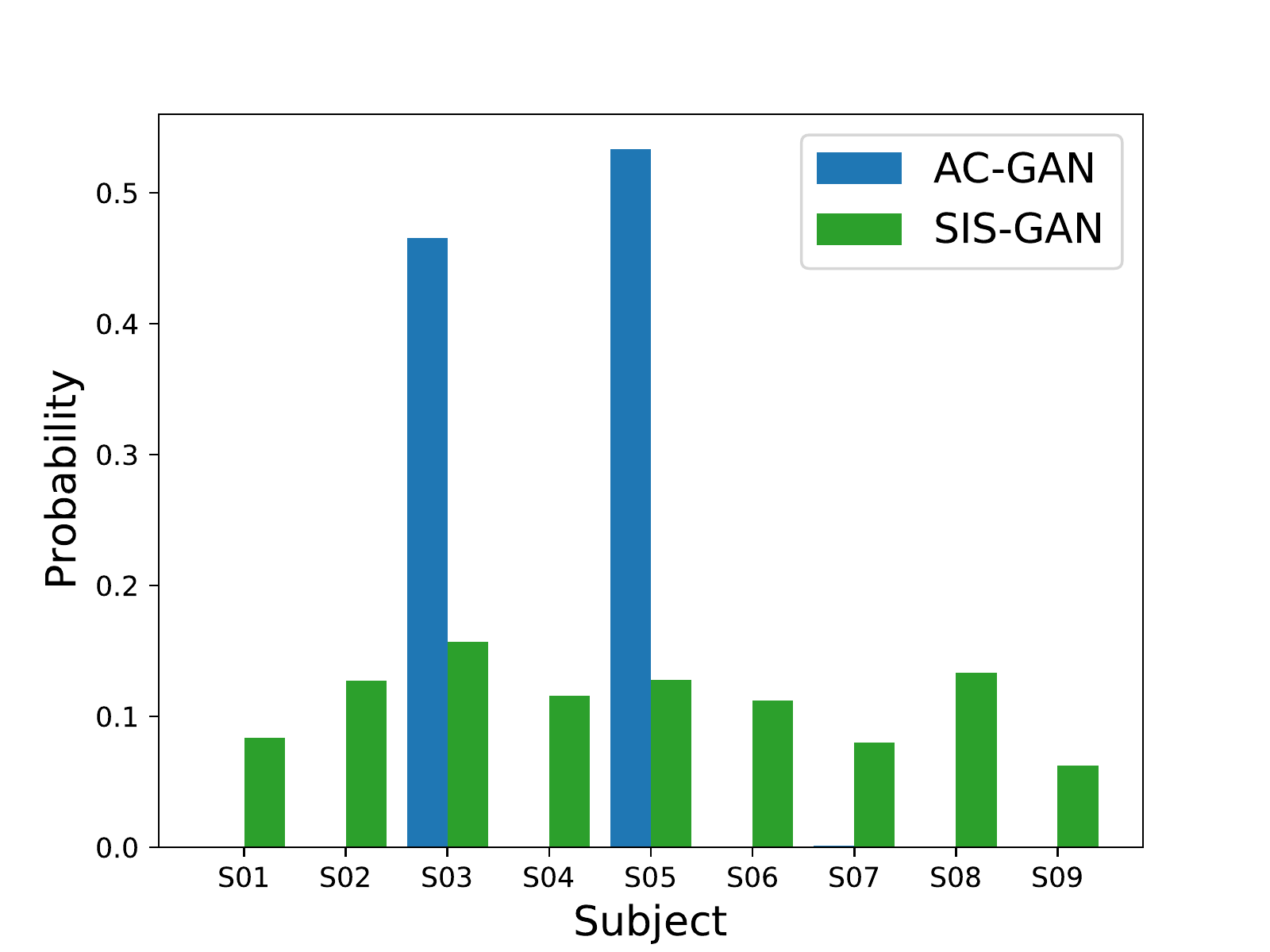} 
  \vskip -5pt
  \caption{Softmax probability values taken from the pre-trained subject-biometric classification network for the generated data. A low consistent value distributed across all subjects is best as it indicates the subject-biometric classification network is unable to find discriminative features.}
  \vskip -20pt
  \label{fig:sub_class}
\end{figure}

To further explore our approach, we investigate the properties of the subject-invariant synthetic data generated by SIS-GAN by visualising the softmax probability assigned to the generated data by the pre-trained subject-biometric classification network. Here, we hope to observe that the pre-trained network is unable to find any features in the generated data that can be used to classify the subject. The results from this experiment are displayed in Figure \ref{fig:sub_class}, where data generated by both the AC-GAN and SIS-GAN approaches are passed through the pre-trained subject network and the resulting softmax probability values are recorded. The figure shows that data generated by AC-GAN is often classified as belonging to either subject S03 or S04, meaning that strong subject-specific features are clearly still present in the data. However, the data generated by SIS-GAN has been assigned a low probability distributed across all nine subjects, indicating that the subject-biometric classification network is unable to find discriminative subject features in the data. This result further indicates that SIS-GAN produces subject invariant data  still containing the desired SSVEP frequency. 

\section{Conclusion}

Training learning-based BCI systems capable of generalising across subjects is currently an open research problem. In practice, many online BCI tasks require calibration to be performed for each new subject before an accurate prediction can be made. In this work, we have explored the possibility of generating new synthetic SSVEP-based dry-EEG data, via neural-based generative models, to help aid with this problem and move towards the removal of the calibration stage altogether. We have explored a novel approach, SIS-GAN to create subject-invariant data by attempting to remove subject-specific features, whilst preserving the SSVEP frequencies we are interested in. Our experimental results demonstrate the efficacy of the synthetically-generated data in improving the performance of a downstream SSVEP frequency classification model. In fact, by training only on synthetic data, we are able to improve the unseen subject generalisation when performing zero-calibration classification by up to 16\% for cross-task classification. For future work, incorporating this zero-calibration BCI application into a real-time teleoperation system can provide excellent testing opportunities. Additionally, the subject invariance within the synthetic data can be further enhanced via a more targeted objective function utilising something analogous to the inception score specifically aimed at reducing subject-specific features within the data.




\bibliographystyle{IEEEtran}
\bibliography{ref}

\end{document}